# A THEORY OF CHAORDIC ECONOMICS: HOW ARTIFICIAL INTELLIGENCE AND BLOCKCHAIN TRANSFORM BUSINESSES, ECONOMIES, AND SOCIETIES

HORST TREIBLMAIER

*School of International Management, Modul University Vienna,*
*Am Kahlenberg 1, Vienna, 1190 Vienna, Austria*
*horst.treiblmaier@modul.ac.at*
*https://www.researchgate.net/profile/Horst-Treiblmaier*

Dee Hock, the founder of Visa, coined the term "chaordic" to describe simultaneously chaotic and ordered systems. Based on his reasoning, we introduce the Theory of Chaordic Economics to explain how economic systems are transformed by two disruptive technologies: Artificial Intelligence and Blockchain. Artificial intelligence can generate novel output through algorithmic yet rather unpredictable processes. Blockchain creates deterministic results without central authorities and relies on elaborated protocols that prescribe how consensus can be reached within a network of peers. The amalgamation of chaos and order produces chaordic economic systems and can yield hitherto unthinkable economic structures.

*Keywords*: Chaordic Economics; Blockchain; Artificial Intelligence; Theory Development.

## 1. Introduction

For better or worse, the evolution of humankind is invariably linked to the governance structures it creates to organize economic and social endeavors. This pertains to the exchange of goods and services (mainly in the private sector) and the mechanisms to establish organizational and societal governance (in the public and private sectors). The emerging structures are fundamentally determined by the underlying technological means, especially in relation to the flow and processing of information needed to understand what is happening in remote locations and act accordingly. Due to the lack of sophisticated information technologies, relatively simple social hierarchies prevailed in the Stone Age, and were restricted to small and controllable areas. From 550 BC onward, great empires emerged (e.g., in Persia, China, Rome, Byzantium, India), sustained by a combination of military force, economic incentives, cultural integration, and administrative innovations. As these empires have disappeared, others have followed, and today's world is characterized by the political and economic dominance of a relatively small number of national states, only a handful of which control the world's economy together with multinational corporations governed by intricate mechanisms that have evolved over centuries.

In the 1960s, motivated by his ongoing quest to shape organizations from how they are to how they ought to be, Dee Hock launched an international payment system, later branded Visa, as a deliberately decentralized organization. He was also the one who coined the term "chaordic" since he concluded that traditional approaches would not suffice to facilitate the growth needed for his envisioned worldwide organization (Hock 2005). "Chaordic" is a portmanteau illustrating the need to balance flexibility or unpredictability (chaos)[*] with clear structure (order) and to allow an organization to innovate and thrive while remaining resilient in turbulent environments.

Academic research has highlighted the relevance of chaordic systems and their implications for organizational and social structures. For example, van Eijnatten and Putnik (2004) propose that learning organizations might evolve into "chaordic enterprises" capable of self-organization and transformative change. Edwards (2014) evaluates and summarizes the literature on chaordic systems thinking applied to organizational transformation and identifies several so-called metatheoretical lenses that can be used in chaordic systems thinking, including connectivity, emergence, indeterminacy, open systems, agency-communion, dialogue, holarchy, dissipation and consciousness, and the individual vs. collective lens.

However, academia still lacks an overarching theory on chaordic systems that scrutinizes the impact of disruptive technologies and, importantly, provides a theoretical explanation for how technological transformation allows the reorganization of existing structures, potentially fostering economic growth and welfare. This is understandable since Dee Hock's original ideas were driven by practical needs rather than academic inquiry. To better understand the implications of modern technologies for economic and societal change and to create awareness of forthcoming transformations, we postulate that it is the available tools that initially determine the capabilities and that the capabilities, in turn, determine the structures. We consider information technology, specifically tools for the collection, storage, processing, transfer, and use of information, as an essential driver of change, and consequently claim that advances in information technology ultimately transform structures on various levels by creating new possibilities. In a day and age in which light travels through fiber optic cables at about 200,000 km per second, decentralized ledgers facilitate consensus and create trust among entities who do not know each other, and AI-powered systems take mere seconds to process more information than a human could contemplate in a lifetime; it is time to question whether the structures and processes that worked considerably well during the industrial revolution[†] are still the best available means to organize work and life in the 21st century.

In this chapter, we propose and logically justify the Theory of Chaordic Economics that specifically considers two important and presumably complementary technologies: Artificial Intelligence (AI) and Blockchain. We argue that both can lead to unpredictable and chaotic outcomes, but can also help to create orderly and structured systems. Together, they can provide the basis for a new economic order and yield hitherto impossible structures for organizations, economies, and societies.

---

[*] Various definitions of chaos and order exist. In this chapter, we take a rather broad approach and view chaos as unpredictability and order as predictability, leading to unstructured and structured outcomes, respectively.
[†] They have worked well to foster innovation and economic progress on a national level in industrialized nations using standard economic metrics. If the ecological consequences and the implications for the subjugated nations and the working classes were considered, the final assessment would be a different one.

## 2. Theories on Chaos and Order

Numerous existing theories can be applied to analyze, explain, and predict the emergence of chaos vs. order and identify relevant driving factors. In this chapter, we focus on a handful of rather formal mathematical theories and others that strive to explain economic outcomes. To start with, Information Theory is rooted in the seminal work by Shannon (1948), in which he mathematically formalized the core elements of communication and developed the concept of information entropy, expressed as the unpredictability of information content. From an organizational, economic, and societal perspective, entropy can be used to quantify the amount of uncertainty in a system, with organizations operating somewhere on a spectrum ranging from predictable (low entropy) to unpredictable (high entropy) environments. Information Theory also elucidates the interplay between information flows and organizational structures, with hierarchical structures generally featuring clearly defined flows, while emergent structures are characterized by rather unstructured, decentralized interactions. From the viewpoint of Information Theory, self-organization can happen through unplanned interactions and patterns, allowing structure and order to emerge.

Chaos Theory emerged as a branch of mathematics that investigates how minor variations in initial states lead to vastly different outcomes. Empirically discovered through meteorology (Lorenz 1963), it has found widespread application in fields such as mathematics, physics, biology, medicine, computer science, the social sciences, and economics. At the core of this theory is the assumption that complex systems that appear to be chaotic may have an underlying order and can even have a deterministic foundation. In this regard, deterministic chaos describes a situation in which order (i.e., deterministic rules) and chaos (i.e., unpredictable outcomes) co-exist. From an organizational, economic, and social perspective, this implies that even seemingly ordered systems might contain the potential for unpredictable change.

Cybernetics is a transdisciplinary field that investigates systems in which outputs become inputs, resulting in feedback loops that can regulate the overall system, adapt to changes, and help to achieve desired goals. It originates in the work of Wiener (1948), who combined ideas from mathematics, biology, and engineering to design systems of control and communication. From a cybernetic perspective, systems become chaotic when control mechanisms are inadequately designed or fail, as is the case with insufficient governance or inappropriate policies. Conversely, it is also possible that order emerges from chaos, as happens when markets reach a new equilibrium or communities self-regulate during times of crises. In short, communication, feedback, and control can help to stabilize systems through efficient information flows, dynamic adjustment, and coordinated regulation.

The focus of Game Theory lies in the analysis of strategic interactions among agents, which can be individuals, organizations, or other economic entities. These agents make decisions in scenarios that can be modeled as games in which the choices and outcomes depend not only on their own actions but also on the actions of others, such that strategic action often requires consideration of mutual expectations. Similar to the theories discussed above, the origins of game theory can be found in mathematics, with the seminal works of John von Neumann and Oskar Morgenstern laying the foundation. The initial ideas were later refined by Nash (1951), who extended von Neumann's work to non-cooperative games by introducing the concept of the Nash Equilibrium. He provided a theoretical framework to understand the outcome of games in which players make decisions independently. In a nutshell, it is the nature of the interactions that leads either to conflict, which often implies chaos, or order enabled by structured cooperation.

At the heart of Network Theory is the study of interactions between networked nodes, which could be individuals or organizations, and how information flows among them. Its origins can be traced back to the works of the mathematician Leonard Euler, who laid the foundation for Graph Theory in the 18$^{th}$ century and inspired significant work in that area, including the Erdős-Rényi model of random graphs, which fostered understanding of how networks grow and evolve. Albert and Barabási (2002) introduced the concept of scale-free networks and the preferential attachment model, in which they illustrate that many real-world networks, such as the internet and social networks, follow a power-law degree distribution, with only a few highly connected nodes. In such networks, local interactions can serve both as the foundation of order and a source of instability. Further developments include Economic network theory (Swan 2019), which indicates that the widespread adoption of distributed ledgers can help mitigate systemic risks in financial networks.

When it comes to the emergent organization of the economic sector, Coase (1937) posed the fundamental question of why firms exist in the first place. He concluded that the so-called transaction costs (i.e., expenses for search and information, negotiations, enforcement, and compliance) ultimately determine a firm's boundaries. Simply put, firms will grow if conducting transactions internally is cheaper than on the open market. Environments characterized by uncertainty and complexity tend to yield hierarchies that can cope with outside conditions and foster hierarchical structures for control.

Austrian Economics represents a school of thought that emphasizes the role of individual decision-making, subjective value, and the importance of free markets to foster economic coordination and growth (von Mises 1949). The core concepts were developed and refined by numerous economists, including Carl Menger, Ludwig von Mises, Friedrich Hayek, and Murray Rothbard. They postulate that it is primarily interference with free market mechanisms that leads to the emergence of chaos since the natural flow of information, expressed by prices in voluntary exchanges, becomes distorted. Consequently, they routinely blame governmental interventions for causing such distortions. On the other hand, they claim that economic order can be established by allowing decentralized decision-making and letting individuals pursue their self-interests.

In stark contrast to the Austrian Economists, John Maynard Keynes (1936) concentrates on the roles of aggregate demand fluctuations, general uncertainty, and wrong policy decisions in creating chaotic economic environments. He postulates that market inefficiencies can be corrected and stable economic conditions ensured by focusing on the management of demand through controlled governmental interventions, including counter-cyclical policies using monetary and fiscal measures.

Finally, proponents of Behavioral Economics focus on the understanding of how humans make economic decisions. They deviate from the underlying assumptions of unbounded rationality and pure self-interest, by identifying cognitive biases, emotions, and social influences as important determinants of decision-making processes. From an individual's perspective, gains and losses are not evaluated objectively but rather compared to a subjective reference point (Kahneman and Tversky 1979). According to this theory, chaos in economic systems can be explained by (objectively) irrational human behavior leading to unpredictable, inconsistent, or destabilizing outcomes. These behaviors, aggregated across individuals or institutions, can disrupt market equilibria, amplify volatility, and create systemic instability. While individual behaviors may appear irrational when viewed in isolation, behavioral economics demonstrates that these behaviors are often systematic and predictable. When aggregated, they form structured patterns that explain market trends, policy outcomes, and social dynamics.

In Table 1, we summarize these different perspectives on the drivers of order and chaos, focusing on the important role of information in shaping the emergent organizational, economic, and societal structures.

Table 1. Theoretical perspectives on chaos and order.

| Theory / Framework | Chaos | Order |
|---|---|---|
| Information Theory | Information entropy as disorder | Self-organization through unplanned interactions |
| Chaos Theory | Ordered systems can exhibit unstable and unpredictable behavior | Identification of conditions that yield ordered systems |
| Cybernetics | Inadequate or broken control mechanisms | Stabilization through communication, feedback and control |
| Game Theory | Cooperation yields order | Conflict yields chaos |
| Network Theory | Local interactions as a predictor for global disorder | Local interactions as the foundation of self-organization and hierarchy |
| Transaction Cost Economics | Uncertainty and complexity | Hierarchical structures for internal coordination and cost control |
| Austrian Economics | Imbalances caused by external distortions and disruptions of natural market processes | Decentralized decision-making yields ordered economic activity |
| Keynesianism | Uncertainty and inadequate responses can increase instability | Creation of order through strategic interventions |
| Behavioral Economics | Individual irrational behavior | Aggregate effects of irrational behavior can create predictable structures |

## 3. Artificial Intelligence and Blockchain as Drivers of Order and Chaos

Russell and Norvig (1995) understand Artificial Intelligence as the study of rational agents that optimize the expected outcome of their actions, classifying the concept into four categories: (1) thinking humanly, (2) thinking rationally, (3) acting humanly, and (4) acting rationally. The first and the third categories follow a human approach, whereas the second and fourth follow a rationalist approach. Blockchain is defined as a "digital, decentralized and distributed ledger in which transactions are logged and added in chronological order with the goal of creating permanent and tamperproof records" (Treiblmaier 2018, 547).

Academic researchers have already recognized the important interplay of Blockchain and Artificial Intelligence. For example, Adel, Elhakeem and Marzouk (2023) analyzed 2,615 peer-reviewed journal articles published between 2017 and 2023 to identify eight emerging themes in which Artificial Intelligence and Blockchain work together. Among others, these include machine learning applications, price prediction of digital currencies, federated learning, industry applications (e.g., in supply chain management), reinforcement learning, and Industry 4.0. Enhancing Blockchain through Artificial Intelligence can yield better security (e.g., through anomaly and fraud detection), improve smart contracts (e.g., code creation and bug detection), improve consensus mechanisms, and optimize auctions and smart grids (Ressi et al. 2024). Contrariwise, Blockchain can serve as an enabler for autonomous Artificial Intelligence in digital native economic and financial institutions by allowing Artificial Intelligence agents to autonomously interact with blockchain-based institutions through the use of private keys (Nguyen Thanh, Son, and Vo 2024). Blockchain can also be used to verify knowledge that is generated through Artificial Intelligence. Decentralized knowledge graphs can be used to address challenges related to biases, hallucinations, intellectual property rights, and data ownership (TraceLabs 2024)

In the following sections, we use the theoretical angles introduced above to investigate how these different yet complementary technologies can impact business, the economy, and society. We start with a brief investigation of how they impact information flows, processing, and entropy. We then focus on topics related to organizational structure and processes before, finally, investigating broader implications related to markets, economies, and societies. These perspectives were conceptually derived from a compilation of the main aspects covered in the aforementioned theories.

### 3.1. *Artificial Intelligence*

Artificial Intelligence exhibits features that can contribute to chaos (i.e., unpredictability) and order (i.e., predictability). It is especially generative Artificial Intelligence that is capable of producing unpredictable output through complex processes. However, behind the apparent chaos lie well-defined algorithms and mathematical models. From an information perspective, Artificial Intelligence can potentially enhance chaos through the creation and propagation of misinformation, which, from the viewpoint of entropy, increases uncertainty and noise in the available information. However, the application of Artificial Intelligence can also have the opposite effects on information, through efficient filtering, fact-checking, and structuring, as well as enhancing the flow of information through constant monitoring and analysis. In this context, pattern recognition and compression help to reduce the amount of information entropy.

When it comes to decision-making, Artificial Intelligence can yield chaos by amplifying existing biases and generating unpredictable and unverifiable results. Contrariwise, it can also be argued that the decisions taken by Artificial Intelligence are driven by data rather than potentially irrational human reasoning and thus help to create order based on rationality. From the perspective of structure and organization, Artificial Intelligence can create chaos when the emerging structures are misaligned with systemic or human needs, which can happen if the restructuring is done without taking into account important goals (humanitarian or organizational) or if the tools are implemented without proper integration. Order can be created by optimizing processes, streamlining tasks through automation, and providing support for decision making. Artificial Intelligence can also drive chaos in control mechanisms if it increases the unpredictability of control, reduces the influence of humans, or poses a risk for malicious manipulation. Simultaneously, it can lead to order through the automation of control mechanisms and potentially yield more equitable decisions. From the viewpoint of centralization versus decentralization, Artificial Intelligence can create chaotic situations by suggesting conflicting priorities and power imbalances, but it can also help to align interests between central authorities and local autonomies. Emerging situations can lead to either cooperation or conflict, which can be exacerbated through the creation of information asymmetries and misinformation or mitigated through the improvement of collaboration.

On a market level, the introduction of Artificial Intelligence can lead to a destabilization of existing systems, for example through the creation and spreading of misinformation and an amplification of risks. On the positive side, algorithms might be capable of allocating scarce resources more efficiently and stabilizing markets. Consequently, emerging economies can be either plagued by a disruption of the existing workforce and increasing economic inequalities or aided by a more efficient allocation of resources and the streamlining of workflows. Finally, social structures might be plunged into chaos through restructuring which exacerbates inequalities, or, conversely, the application of Artificial Intelligence may yield more efficient and fair governance processes and foster inclusivity. In Table 2, we summarize the different perspectives on how Artificial Intelligence can contribute to the emergence of chaos and order from different angles.

Table 2. Artificial Intelligence as a driver of chaos and order.

|  | Chaos | Order |
| --- | --- | --- |
| Information processing | Creation of misinformation | Filtering, fact-checking, and structuring |
| Information flow | Propagation of misinformation | Monitoring and analyzing in real-time |
| Information entropy | Increasing uncertainty and noise | Pattern recognition and information compression |
| Decision-making processes | Amplification of biases and unpredictability | Data-driven and organized decisions |
| Organizational structures | Misalignment with systemic and human needs | Process optimization, task automation, and decision support |
| Control mechanisms | Increasing unpredictability, removing human oversight or manipulation | Task automation and fairness |
| Centralization vs. decentralization | Conflicting priorities and the creation of power imbalances | Aligning central authority with the interests of local autonomies |
| Cooperation vs. conflict | Creating information asymmetries and misinformation | Improved collaboration |
| Market processes | Destabilization of existing systems and amplification of risks | Improving resource allocation and market stabilization |
| Economic activities | Disruption of the workforce and economic inequality | Efficient resource allocation and streamlining workflows |
| Social structures | Changing power structures and the creation of inequality | Supporting governance and decision-making and inclusivity |

### 3.2. *Blockchain*

Blockchain can induce unpredictability in information processing through fragmented systems that lack interoperability and the complexity of smart contract code. It can also promote order since the execution of these contracts is deterministic, and the transactions can be traced on a transparent ledger. In this respect, the "Internet of Contracts" can completely transform the way in which legal contracts are created, enforced and regulated (Noto La Diega 2022). Information flows might be negatively impacted by unregulated dissemination and the amplification of feedback loops caused by real-time transaction processing and feedback. In chaordic and non-linear systems, such loops might result in unpredictable dynamics. However, the traceability of the transactions also stabilizes the system and improves transparency and trust. From an information entropy perspective, Blockchain can enhance chaos, defined as unpredictability, by increasing data redundancy (a defining characteristic of distributed ledgers), decentralization, introducing inconsistencies caused by forks, and potentially also information silos due to the proliferation of different blockchains. At the same time, distributed ledgers reduce the amount of entropy through standardized protocols, improved reliability since previous transactions cannot be altered, and increased security through the application of cryptographic procedures.

When it comes to decision-making processes in Chaordic Economics, Blockchain can induce chaos through its reliance on decentralized authorities, which implies interactions between entities with potentially conflicting goals. Depending on the decision-making process, certain groups of actors may be able to take over by acquiring a significant amount of governance tokens, with or without malicious intentions. At the same time, achieving consensus within a network of peers is transparent and based on clearly specified mechanisms such as Proof-of-Work or Proof-of-Stake. Compliance can be enforced via rewards, and decisions are final and can be trusted. The effects on existing structures of transitioning to a blockchain-based system can also be twofold: While it is possible that decentralization of control and power raises conflicts between stakeholders and makes it hard to govern, the rules for governance also become more transparent and the relationships between the various stakeholders are clarified.

In terms of control mechanisms, the introduction of blockchain-based governance can increase complexity due to the decentralization of authority and weaken existing control structures. This might be mitigated or compensated by the introduction of clear incentives for the creation of orderly structures, the provision of mechanisms that help to resolve conflicts, and the easy enforcement of accountability. In terms of (de)centralization, Blockchain can foster chaos through conflicts between centralized and decentralized authorities, which might be aggravated by the emergence of centralized structures within decentralized systems (e.g., disproportionate mining power in PoW systems or unequal token distribution in PoS blockchains). However, such systems can also exhibit greater flexibility, which strengthens their resilience in complex and ever-changing environments. The introduction of Decentralized Autonomous Organizations (DAOs) can also disrupt well-established conflict resolution mechanisms, thus leading to increased chaos, expressed, for example, by the forking of blockchains. The same tools can also have the opposite effect and create a stable foundation for cooperation by building trust that is based on the transparency of protocols and transactions, which, in turn, leads to a better alignment of goals.

When it comes to markets, Blockchain can increase chaos by leading to increased fragmentation that splits markets into competing ecosystems. The "tokenization of everything" can potentially lead to a hyper-financialization that yields speculative bubbles and increases volatility. Simultaneously, stabilizing effects can be expected from the transparency of Blockchain, which not only allows transactions to be easily audited but can also increase overall market efficiency. On an economic level, well-established structures might be displaced by rather untested economic models, which can increase speculation and volatility and make regulation and oversight more complex. On the other hand, ordered systems might emerge that automatically enforce regulatory compliance and promote the optimized allocation of resources to provide sufficient security for long-term planning.

The introduction of Blockchain can have chaotic effects on social structures and lead to the erosion of well-established hierarchies, the introduction of (new) inequalities, and an overreliance on protocols that have not been adequately tested on a large scale. Stabilizing effects on societal structures can be expected from introducing protocols that ensure the fair distribution of value, improve the overall accountability of decision makers (which can be protocols), and streamline the resolution of conflicts. Table 3 summarizes the roles of Blockchain as a driver of chaos and order from numerous perspectives.

Table 3. Blockchain as a driver of chaos and order.

|  | Chaos | Order |
| --- | --- | --- |
| Information processing | Fragmentation and complexity | Deterministic execution, transparency, and traceability |
| Information flow | Unregulated information dissemination and amplification of feedback loops | Increased transparency and trust |
| Information entropy | Redundancy, uncertainty, and inconsistencies | Standardization, reliability, and security |
| Decision-making processes | Decentralized authority, goal conflicts, and malicious interference | Consensus mechanisms, compliance, and finality |
| Organizational structures | Decentralization of control and power, lack of governance, and stakeholder conflict | Transparent governance and clarification of stakeholder relationships |
| Control mechanisms | Increased complexity, decentralized authority, and loss of traditional control structures | Incentives for order, mechanisms for conflict resolution, and enforcement of accountability |
| Centralization vs. decentralization | Conflict of interests between centralized and decentralized authorities, centralization tendencies | Increased flexibility and resilience |
| Cooperation vs. conflict | Disruption of existing conflict resolution mechanisms | Transparency-based trust and shared goals |
| Market processes | Emergence of fragmented ecosystems and amplification of volatility | Efficient and transparent processed |
| Economic activities | Disruption of well-established structures and proliferation of untested economic models | Predictable economic frameworks, efficient allocation of resources, and built-in regulation |
| Social structures | Erosion of traditional hierarchies, inequality, and overreliance on technology | Fair value distribution, improved accountability and streamlined dispute resolution |

## 4. Chaordic Economics: Blurring Boundaries and Emerging Structures

We introduce the Theory of Chaordic Economics as a theoretical framework to analyze, explain, and predict current and future transformations that affect businesses, economies, and societies. Chaordic Economics is based on the transformative force of disruptive technologies such as Artificial Intelligence and Blockchain and characterizes an adaptive state in which economic systems (including societal structures) operate efficiently and effectively. In the following sections, we first outline the key determinants that characterize the Theory of Chaordic Economics, followed by a high-level visualization of how the underlying technological drivers transform organizational structures, inter-organizational relations and macroeconomic processes, and societal structures.

### 4.1. *Key Characteristics*

We postulate that organizations, economies, and societies that transform into a chaordic state exhibit several critical characteristics that distinguish them from traditional mechanistic and hierarchical structures.

*Self-organization and emergence*

In chaordic systems, various forms of economic transactions and structures emerge from the interactions of numerous individual agents (which can be humans but also machines) rather than from rigid top-down control. This corresponds to the idea of a spontaneous order of free markets but also recognizes the need for guiding principles to prevent destructive states of disorder. Such principles could be encoded in smart contracts.

*Adaptability and resilience*

By nature, chaordic economic systems are designed to be resilient, adaptive, and flexible. Decentralized approaches enable such systems to adapt to disruptions, such as technological change, environmental disasters, or economic shocks. This is achieved through feedback loops, which help the systems develop and adjust their structure to be able to cope with external events.

*Non-linearity*

The complex interplay between chaos and order can yield unpredictable behaviors that expose non-linear outcomes. Feedback loops and tipping points create emerging patterns that cannot be explained by traditional linear models and necessitate broad systemic thinking and an increased focus on adaptability.

*Balance between chaos and order*

Chaordic economies exhibit a balance where structure exists to provide stability, but sufficient flexibility is granted to allow for innovation and adaptability. The extreme endpoints are avoided, recognizing that a rigid and hierarchical order can prevent growth and innovation, while uncontrolled chaos can lead to instability.

*Decentralized control*

Chaordic economics does not rely on central planning or laissez-faire approaches but is based on a balanced middle ground in which general guidelines and principles steer behavior without rigid control. This encourages market-driven open innovation while simultaneously ensuring stability through shared principles, values, and, if needed, regulations.

*Stakeholder Integration*

Chaordic economic systems are open to diverse stakeholders, including individuals, businesses, local communities, and governments. Rather than focusing entirely on competition to foster development, Chaordic Economics emphasizes collaboration and the alignment of potentially diverse interests. From this perspective, economies and societal structures are seen as interconnected ecosystems that evolve together and influence each other.

*Complexity*

Chaordic economics considers the interconnected and rather unpredictable nature of economies. It deviates from reductionist thinking which implies that systems can be best understood if broken down into parts, and acknowledges that economic outcomes result from complex and frequently non-linear interactions. This acknowledges that our world is characterized by increasing technological, economic, and social complexity.

*Sustainability*

An integrative view of sustainability incorporates environmental, economic, and social aspects. In this regard, Chaordic Economics integrates flexible and self-organizing processes that follow the principles of the circular economy while simultaneously considering economic and social disparities. Such an approach favors long-term stability over short-term gains, leveraging feedback loops and emergent behaviors to align economic goals with sustainable practices.

**4.2. Integrative Theoretical Framework**

Fig. 1 depicts the blurring boundaries between organizational structures, economic conditions (shown as inter-organizational relationships and macroeconomic conditions), and societal structures. In a chaordic economic system, the borders between centralization and decentralization, as well as private and public, become permeable, as is shown by the two-headed arrows. Based on our reasoning, as outlined above, the direction of a specific development is hard to predict, but, in general, chaordic systems tend to be more decentralized and autonomous than existing hierarchical structures. Private organizations or DAOs can take over social tasks (e.g., related to governance or voting), and governing structures will evolve into self-organizing autonomous systems.

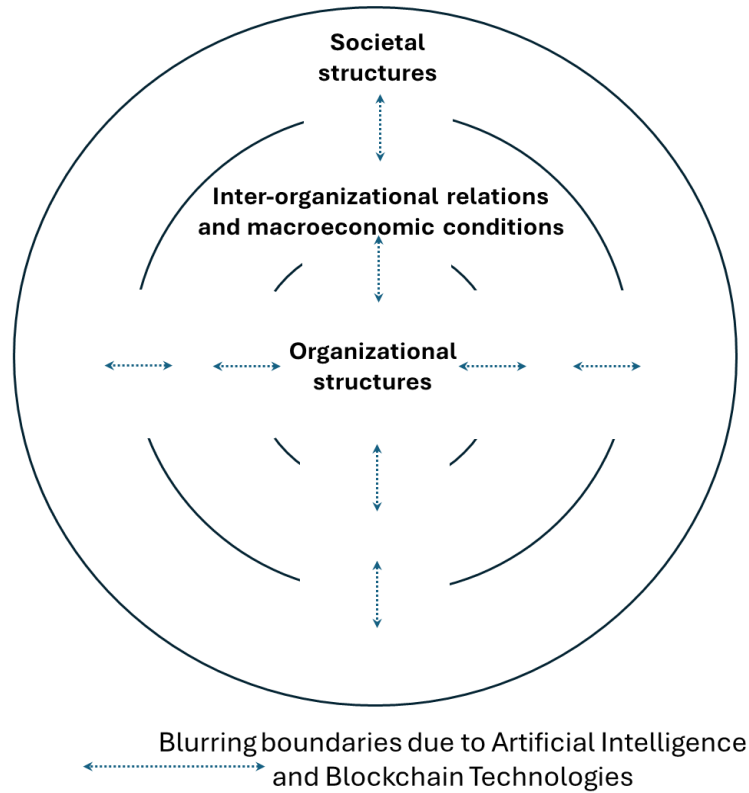

Fig. 1. Blurring of Boundaries in Chaordic Economics.

## 5. Discussion

Existing economic and social structures, established over centuries under fundamentally different conditions than today, no longer pose the most effective and efficient ways to manage organizations, steer economies, and govern societies. In this chapter, we postulate that if the underlying conditions and the technological means have changed, it is reasonable to assume that new structures and processes will emerge that will alter all levels of economic and social activity. In this regard, we build on seminal literature and emphasize the need for systematic thinking rather than analyzing individual parts in isolation (Meadows 2008). Our reasoning is based on the assumption that information is the key to governing organizations (Laudon and Laudon 2020) and societies (Beniger 1986). As a starting point, we use numerous existing theories to gain a better understanding of how chaos and order emerge in different kinds of systems. In addition to these classic theories, new approaches, such as Smart Network Field Theory (Swan and dos Santos 2020), have already indicated that novel technologies, including Blockchain and Artificial Intelligence, can yield disruptive and autonomously operating networks that lead to new economic equilibria and how such developments can be explained from the viewpoint of the natural sciences.

The proposed Theory of Chaordic Economics is a grand theory outlining broad and fundamental aspects that need to be further broken down into smaller components and frameworks, labeled as middle-range theories, which lend themselves to rigorous academic inquiry. This will help to analyze, explain, and predict how systems emerge, change, and adapt that are partially chaotic in that they exhibit adaptive and decentralized structures and are partially ordered since they are determined and controlled by underlying protocols and algorithms following pre-determined rules. Over time, self-organizing networks (e.g., DAOs) emerge that exhibit the core characteristics outlined in this chapter.

Following the classification of Gregor (2006), we suggest theory-based inquiries for the suggested Theory of Chaordic Economics. Table 4 shows exemplary research questions as starting points for rigorous academic research. Ideally, and in line with the multidisciplinary nature of Chaordic Economics, these questions are tackled by multidisciplinary teams comprising insights from economics, sociology, computer science, information systems, systems science, and business administration. These questions are meant to analyze existing systems transitioning into a chaordic state, explain the occurrence of specific events and transformations, and predict future development based on technological development. Additionally, the Theory of Chaordic Economics can be used as a guiding framework (i.e., a worldview) to design and implement systems that exhibit desired properties and to steer future developments.

Table 4. Research questions for the Theory of Chaordic Economics.

| Theory Type | Questions |
| --- | --- |
| Analyze and explore | What are the core characteristics of Chaordic Economics? |
| | How are the core characteristics of Chaordic Economics related to each other? |
| | How do external factors (e.g., technology, regulation, culture) shape the emergence of Chaordic Economics? |
| | What are the fundamental values underlying the Theory of Chaordic Economics? |
| Explanation | Why does a specific type of behavior emerge in Chaordic Economics? |
| | How can emergent behavior be modeled in Chaordic Economics? |
| | How do businesses adapt to Chaordic Economics? |
| | What role do feedback loops play in the shaping of Chaordic Economics? |
| Prediction | How will chaordic economic systems evolve if the environmental conditions change? |
| | What factors predict the success of failure of Chaordic Economics? |
| Design and Action | How can systems be designed that follow the principles of Chaordic Economics? |
| | How can systems be deployed that follow the principles of Chaordic Economics? |
| | How can systems be controlled that follow the principles of Chaordic Economics? |

From a practitioner's point of view, the Theory of Chaordic Economics provides multiple angles for managers and policymakers to investigate and better understand current economic and social transformations and to act accordingly. A comprehensive systemic perspective allows them to simultaneously consider disruptive forces and acknowledge the complex behavior of emerging systems. Allowing better systems to emerge might imply that policymakers relinquish control, which, presumably, will be a huge mental hurdle for those in power. In this regard, it has to be acknowledged that the Theory of Chaordic Economics has a strong normative component, originating in Dee Hock's initial quest to find out how organizations *ought* to be. Furthermore, it is crucial to emphasize that all important components must be considered simultaneously, and the neglect of fundamental underlying values (e.g., the right to privacy) cannot be justified by exclusively focusing on parts of the whole, such as economic efficiency.

## 6. Conclusion, Limitations, and Further Research

In this chapter, we propose the Theory of Chaordic Economics, which is based on the fundamental assumption that the flow, processing, and use of information are changing and that this development will significantly affect businesses, economies, and societies. Global economic and societal systems are not static but constantly evolving, and technologies are major enablers and drivers in this development. We postulate that this will lead to the emergence of chaordic systems, which can adjust to external changes through adaptation and self-organization while concurrently maintaining stability and structure. Chaordic Economics helps to establish adaptive businesses, economies, and societies that can thrive in constantly changing environments and can adjust to technological, market, and political changes.

Despite promising economic and societal gains, this transformation will presumably face powerful and stubborn opponents. Individuals in command will strive to control the uncontrollable, and technology-enhanced supervision capabilities can support them to stay in power. The devastating effects of useless administration and compulsive control are by no means new phenomena and have been explained decades ago pointedly and humorously by Parkinson (1958). He described how strict hierarchies cause a lack of innovation and how overburdened bureaucracies stifle innovation and employees' motivation by creating useless work and control structures with no purpose other than bolstering the egos of those in control. Parkinson used the example of the British Royal Navy, in which bureaucracy increased over time even though the actual fleet size and the number of sailors decreased. His warnings are even more relevant during times when mindless Excel sheets can serve no purpose other than cementing existing power structures. In this regard, we believe that chaordic systems represent a development that has long begun and will be hard to stop.

Our conceptual research faces a couple of limitations that present a roadmap for future research. First, we acknowledge that technologies other than Artificial Intelligence and Blockchain might also play an important role in shaping future developments. For example, the Internet of Things, robotics, or quantum computing might also serve as disruptive technological drivers. Second, numerous criticisms and challenges exist regarding Artificial Intelligence and Blockchain, both individually and in their interplay. For example, both technologies can be notoriously energy-intensive (in the case of Blockchain, this pertains mainly to the use of Proof-of-Work as a consensus mechanism), but there are also other issues such as scalability, interoperability, security, privacy, and potentially also regulatory issues, calling for extensive and structured future investigations (Treiblmaier et al. 2021).

In summary, we suggest that the Theory of Chaordic Economics presents a promising starting point for researchers and practitioners who wish to better understand existing systems, their transformations, and the potential to shape them in a way that fosters humankind rather than staying merely outside observers of these developments.

**Index**